\documentclass[12pt,epsf,showkeys]{article}

\usepackage{graphicx,float}
\usepackage{mathrsfs,array,multirow}
\usepackage{amstext}
\usepackage{subfigure}
\usepackage{bm,latexsym,amsmath,amsfonts,amssymb}
\usepackage[usenames,dvipsnames]{color}
\usepackage{color}
\usepackage[colorlinks=true,linkcolor=blue]{hyperref}
\usepackage{soul}
\usepackage{epsfig}

\newcommand{\be}{\begin{equation}}

\newcommand{\ee}{\end{equation}}

\newcommand{\ba}{\begin{array}}

\newcommand{\ea}{\end{array}}

\begin{document}
\begin{titlepage}
\vspace{.5in}
\begin{flushright}
\end{flushright}
\vspace{0.5cm}

\begin{center}
{\Large\bf Charged traversable wormholes: charge without charge}\\
\vspace{.4in}

  {$\mbox{Hyeong-Chan\,\, Kim}^{\P}$}\footnote{\it email: hckim@ut.ac.kr},\,\,
  {$\mbox{Sung-Won \,\, Kim}^{\ddag}$}\footnote{\it email: sungwon@ewha.ac.kr},\,\,
  {$\mbox{Bum-Hoon \,\, Lee}^{\S\dag}$}\footnote{\it email: bhl@sogang.ac.kr},\,\,
  {$\mbox{Wonwoo \,\, Lee}^{\S}$}\footnote{\it email: warrior@sogang.ac.kr},\,\, \\

\vspace{.3in}

{\small \S \it Center for Quantum Spacetime, Sogang University, Seoul 04107, Korea}\\
{\small \dag \it Department of Physics, Sogang University, Seoul 04107, Korea}\\
{\small \P \it School of Liberal Arts and Sciences, Korea National University of Transportation, Chungju 27469, Korea}\\
{\small \ddag \it Ewha Womans University, Seoul 03760, Korea}\\

\vspace{.3in}

\vspace{.5in}
\end{center}
\begin{center}
{\large\bf Abstract}
\end{center}
\begin{center}
\begin{minipage}{4.75in}

{\small \,\,\,\,  We present and investigate charged wormhole solutions of the Einstein-Maxwell equations supported by anisotropic matter fields,
with the purpose of establishing their physical plausibility as traversable wormholes.
To this end, we examine the flare-out condition and evaluate tidal forces to confirm their traversability.
We also analyze light deflection around these wormholes to provide observational implications.
Additionally, we attempt to construct rotating generalizations of the solutions by applying and modifying the Newman-Janis algorithm.
Our results suggest that the obtained geometries offer a concrete realization of the concept of  ``charge without charge".
 }
\end{minipage}
\end{center}
\end{titlepage}

\newpage

\section{Introduction \label{sec1}}

\quad
The Einstein theory of gravitation~\cite{Einstein:1916vd} has undergone extensive verification through numerous experimental tests over the past century~\cite{LIGOScientific:2016aoc, LIGOScientific:2017vwq, EventHorizonTelescope:2019dse, EventHorizonTelescope:2022wkp},
reaching a mature level of acceptance as a theory of gravitation.
However, the absence of a clear explanation for dark matter
and dark energy is igniting interest in modified theories of gravitation.
Additionally, the ongoing need for precise descriptions of astrophysical phenomena continues to propel research
into finding and analyzing various solutions that incorporate matter fields beyond the vacuum solution.

Among the most intriguing concepts arising from the theory of gravitation is the wormhole solution~\cite{Morris:1988cz},
which theoretically facilitates time travel~\cite{Morris:1988tu, Gott:1990zr, Novikov:1989sd, Shoshany:2019yoz}.
This fascinating notion has captivated scientists and the general public,
making it a compelling subject for exploration in academic study, literature, and movies~\cite{James:2015ima}.

The study of traversable wormholes is gaining increased attention for this reason.
It is also essential to assess the feasibility of travel by evaluating whether the tidal forces
experienced by the traveler (or spacecraft) near the wormhole's throat are comparable to Earth's gravity~\cite{Morris:1988cz}.
This tidal effect can be understood through the components of the Riemann tensor in general relativity.
Additionally, we should investigate the flare-out condition and how the geometry of a wormhole diverges from that of a black hole.
The flare-out condition is a crucial geometric criterion that a wormhole's structure must satisfy.
A geometry that adheres to this condition can be classified as a wormhole.
Importantly, the wormhole geometry must not have any physical singularities.

Einstein and Rosen were among the first to rigorously explore wormhole physics~\cite{Einstein:1935tc},
examining a bridge that connects two identical sheets.
For a detailed history of wormhole physics, please refer to the relevant literature~\cite{Visser:1995cc, Lobo:2007zb}.
The field received a significant boost with the paper by Morris and Thorne~\cite{Morris:1988cz},
leading to extensive research on wormholes across various modified theories of gravitation~\cite{Kim:1995xf, Kim:1997jf, Lobo:2009ip,
Kanti:2011jz, Clement:2015aka, Kim:2016pky, Maldacena:2018gjk, Ovgun:2018xys, Kang:2018izw, Kim:2019elt, Halder:2019akt, Huang:2019arj, Bouhmadi-Lopez:2021zwt, DeFalco:2021ksd, DeFalco:2023twb, Jang:2024nhm} and dimensions~\cite{Agnese:1999fp, Kim:2003zb, Lobo:2007qi, Bak:2018txn, deBoer:2022rbn, Deshpande:2022zfm}.
Additionally, the influence of fermion fields on wormhole geometry has also been the focus of investigation~\cite{Maldacena:2018gjk, Blazquez-Salcedo:2020czn, Konoplya:2021hsm, Danielson:2021aor, Blazquez-Salcedo:2021udn}.

Special matter fields are essential for the construction and maintenance of wormholes.
The matter constituting the energy-momentum tensor of the wormhole geometry violates the null energy condition.
In simpler terms, the matter that supports the wormhole geometry, which adheres to the flare-out condition,
violates the null energy condition as outlined in Einstein's theory, such as phantom energy~\cite{Lobo:2005us}.
In modified theories of gravitation, the effects of modified gravity could
also play a significant role in these special matter fields.

The geometry of a wormhole may serve as a geometric representation of the concept of
``charge without charge", particularly as a solution to the source-free Maxwell equations
in the context of a traversable wormhole.
Misner and Wheeler highlighted the importance of this solution~\cite{Misner:1957mt},
noting that the electric field enters one side of the wormhole and exits the other.
The full realization of this concept would require finding a charged wormhole solution~\cite{Kim:2001ri}.
While the Maxwell field is not essential for constructing a wormhole,
examining its effects on the wormhole geometry and the flare-out condition is intriguing.
The incorporation of matter fields increases the complexity of the equations of motion,
which can potentially be solved either analytically or numerically.
In this paper, we aim to construct a charged wormhole geometry with an anisotropic matter field~\cite{Halder:2019urh, Kim:2019ojs}
and derive Maxwell's solution analytically for this configuration.
Specifically, we focus on the wormhole with the anisotropic matter field obtained in Ref.~\cite{Kim:2019ojs}.

To grasp the distinctions between astrophysical phenomena related to wormholes and black holes in the Universe,
it is essential to comprehend the behavior of light in the vicinity of
wormholes~\cite{Virbhadra:1999nm, Safonova:2001vz, Bozza:2002zj, Nandi:2006ds, Tsukamoto:2012xs, Shaikh:2018oul}.
This understanding begins with an analysis of the geodesics of light.
We then identify the photon sphere and capture cross-section~\cite{Dymnikova1986},
followed by an examination of the light deflection angle around a wormhole.

We are also exploring the geometry associated with rotating objects,
specifically focusing on rotating black holes~\cite{Kim:2019hfp, Kim:2021vlk}.
Since the discovery of the Kerr black hole~\cite{Kerr:1963ud},
significant efforts have been directed toward finding solutions for rotating black holes~\cite{Newman:1965my, Carter:1973rla, Toshmatov:2015npp, Xu:2016jod, Kim:2019hfp, Ali:2021gfc, Ali:2022tdt, Ali:2022omt},
utilizing either the Newman-Janis (NJ) algorithm~\cite{Newman:1965tw, Newman:1965my}
or the method detailed in other reference~\cite{Azreg-Ainou:2014pra}.
Typically, studies of rotating wormholes have relied on the assumption of
a metric ansatz~\cite{Teo:1998dp, Kuhfittig:2003wr, Kleihaus:2014dla, Chew:2018vjp, Clement:2022pjr}.
The NJ algorithm is mathematically well-defined when applied to static geometries,
in which the radial pressure equals the negative of the energy density, $\varepsilon =-p_r$, i.e., the case with $-g_{tt}=g^{rr}$.
However, the energy-momentum tensor that describes the static wormhole geometry does
not satisfy that condition, i.e., $\varepsilon \neq -p_r$.
The first analytic solution for rotating wormhole geometry was derived by Teo~\cite{Teo:1998dp},
who utilized a metric ansatz, while Azreg-Ainou applied the NJ algorithm to the static geometry with the property
$-g_{tt}=g^{rr}$~\cite{Caravelli:2010ff, Azreg-Ainou:2011tkx, Azreg-Ainou:2014aqa} to attain the rotating geometry,
subsequently applying his approach to achieve the rotating wormhole solution~\cite{Azreg-Ainou:2015pda}.
We propose modifications to the NJ algorithm to use it on static geometries
that do not adhere to $\varepsilon+p_r =0$, thus constructing a rotating wormhole geometry.
We aim to extend the NJ algorithm to apply to wormhole geometries characterized by $-g_{tt}\neq g^{rr}$.
To further advance this work, we relax the mathematical rigor slightly and extend the NJ algorithm's application.
We apply this modified approach to construct the geometry of a rotating wormhole
and conduct an analysis of the black hole case for comparison with the wormhole scenario.

This paper is structured as follows:
In Sec.~[\ref{sec2}], we present the charged static wormhole solution,
along with the associated energy density and pressure of the matter fields necessary to support this wormhole geometry.
We analyze in detail the conditions required for our solutions to qualify as traversable wormholes.
We examine the effects of gravitational lensing produced by the wormhole geometry.
In Sec.~[\ref{sec3}], we present our attempts to derive the geometry of rotating wormholes
through the modification of the Newman-Janis algorithm.
In Sec.~[\ref{sec4}], we summarize and discuss our results.
In Appendix~[\ref{app:robh}], we apply the modified NJ algorithm to demonstrate
the geometry of a rotating black hole for comparison with the rotating wormhole geometry.

\section{Charged static wormhole \label{sec2}}
\quad

In this section, we construct traversable, charged, static wormhole solutions.
We begin by solving both the Einstein and Maxwell equations, which are coupled with a matter field.
Following that, we analyze the conditions required for the geometry to qualify as a wormhole.

\subsection{Setup and solution \label{sec2-1}}

\quad

We consider the action~\cite{Kim:2001ri, Halder:2019urh, Kim:2019ojs}
\begin{equation}
I=\int d^4x \sqrt{-g}\Big[\frac{1}{16\pi}(R-F_{\mu\nu}F^{\mu\nu})+{\cal L}_{\rm am}\Big] + I_{\rm b},
\end{equation}
where ${\cal L}_{\rm am}$ describes effective anisotropic matter fields,
$I_{\rm b}$ is the boundary term~\cite{Gibbons:1976ue, Hawking:1995ap},
and we take the Newton constant $G=1$ for simplicity.

Varying the action, we obtain the Einstein equation
\begin{equation}
G_{\mu\nu}=R_{\mu\nu}-\frac{1}{2}R g_{\mu\nu}=8\pi T_{\mu\nu} \,, \label{einsteineq}
\end{equation}
where the stress-energy tensor takes the form
\begin{eqnarray}
T^{\mu\nu}&=&T^{\mu\nu}_{\rm M} +  T^{\mu\nu}_{\rm am} \nonumber \\
&=& \frac{1}{4\pi}(F^{\mu}_{~\alpha}F^{\nu\alpha}-\frac{1}{4}g^{\mu\nu}F_{\alpha\beta}F^{\alpha\beta})+ (\varepsilon_{\rm am} + p_{t {\rm am}}) u^{\mu}u^{\nu} + p_{t{\rm am}} g^{\mu\nu} + (p_{r {\rm am}} - p_{t {\rm am}}) x^{\mu}x^{\nu}  \,, \label{stente}
\end{eqnarray}
where $\varepsilon_{\rm am}$ is the energy density of the anisotropic matter, $u^{\mu}$ is four-velocity, and $x^{\mu}$ is a spacelike unit vector, respectively. The radial and the transverse (lateral) pressures are assumed to be linearly
proportional to the energy density:
\begin{equation}
p_{r {\rm am}}=w_{\rm 1}\varepsilon_{\rm am}\,,~~~ p_{t {\rm am}}=w_{\rm 2}\varepsilon_{\rm am}  \,, \label{eqofstate}
\end{equation}
then stress tensor for the anisotropic matter field in Eq.~(\ref{stente}) can be rewritten as
$T^{\mu}_{\nu {\rm am} }={\rm diag}(-\varepsilon_{\rm am}, w_1 \varepsilon_{\rm am}, w_2 \varepsilon_{\rm am}, w_2 \varepsilon_{\rm am})$.
The source-free Maxwell equations are given by
\begin{equation}
\nabla_{\nu}F^{\mu\nu} = \frac{1}{\sqrt{-g}}[\partial_{\nu}(\sqrt{-g}F^{\mu\nu})] =0 \,. \label{maxwelleq}
\end{equation}

We take the metric for the static spherically symmetric charged wormhole geometry
\begin{equation}
ds^2=  - f(r) dt^2 +  \frac{1}{g(r)} dr^2 + r^2 (d\theta^2 + \sin^2\theta d\psi^2)  \,, \label{stmetric}
\end{equation}
where metric functions $f(r)$ and $g(r)$ denote the redshift function and include the wormhole shape function, respectively.
We consider an asymptotically flat spacetime.

After solving the Maxwell and Einstein equations,
we analyze the conditions for the solution to be a travelable wormhole.

Let us first tackle the Maxwell equation.
For the electrically charged static geometry, $F^{tr}=E^r=\sqrt{\frac{g(r)}{f(r)}} \frac{Q}{r^2}$
satisfy the source-free Maxwell equations (\ref{maxwelleq}).
In the asymptotic rest frame, one could measure the electric field.
That field should be defined in an orthonormal frame, we adopt covariant tetrad shown as
$e^{\hat t}_{\mu} = (\sqrt{f(r)},0,0,0)$, $e^{\hat r}_{\mu} = (0,\frac{1}{\sqrt{g(r)}},0,0)$,
$e^{\hat \theta}_{\mu} = (0,0,r,0)$, and $e^{\hat \psi}_{\mu} = (0,0,0,r\sin\theta)$.
The electric field takes the form of $F^{{\hat a}{\hat b}}=e^{\hat a}_{\mu}e^{\hat b}_{\nu}F^{\mu\nu}$,
which gives $E^{\hat r}=\frac{Q}{r^2}$.

We take the metric functions $f(r)=\left( 1+ \frac{Q^2}{r^2} \right)$
and $g(r) = \left(1 + \frac{Q^2}{r^2} -\frac{b(r)}{r} \right) $ in Ref.~\cite{Kim:2001ri}.
This geometry has the minimum radius at the throat $r=r_{o}$, i.e. $b(r)=b_{o}$ at $r_{o}$ and thus $g(r_{o})=0$.
This gives $b_{o}=(r^2_{o}+Q^2)/r_{o}$,  in which $b_o$ is the physical parameter of the wormhole with given $w_1$.
When $Q$ and $b_o$ are given, there are two locations of the throat, $r_o=\frac{1}{2}(b_o \pm \sqrt{b^2_o - 4Q^2})$.
We take the larger one as
\begin{equation}
\label{worthroat}
r_o=\frac{1}{2}(b_o + \sqrt{b^2_o - 4Q^2}) \,,
\end{equation}
and $b_o > 2Q$.
If $b_o = 2Q$, then $r_o=Q$, which will not satisfy the flare-out condition for later analysis.
When $Q$ vanishes, $b_o=r_o$. The presence of charge $Q$ reduces the size of the wormhole throat.

There exists also a solution that satisfies the Einstein equations in the region
$r\leq r_o$, in which $r_o$ is the smaller one.
This solution will give a geometry describing $0< r \leq r_o$.
At $r=0$, it will have a singularity,
and at $r=r_o$, it will give a geometry connected by a wormhole.
We leave the analysis of this part to future work.

We now consider Einstein equations.
The nonvanishing components of the Einstein tensor are given by
\begin{eqnarray}
G^{t}_{t}&=& - 8\pi \varepsilon = 8\pi( - \varepsilon_{c} - \varepsilon_{\rm am}) = 8\pi(- \varepsilon_{c}) -\frac{b'(r)}{r^2} \,,  \label{Etensor1} \\
G^{r}_{r}&=& 8\pi p_r = 8\pi(- \varepsilon_{c} + w_1 \varepsilon_{\rm am}) = 8\pi( - \varepsilon_{c})
+ \frac{(Q^2-r^2)b(r)}{r^3(Q^2+r^2)} \,,   \label{Etensor2} \\
G^{\theta}_{\theta}&=&G^{\phi}_{\phi}=8\pi p_t = 8\pi (\varepsilon_{c} + w_2 \varepsilon_{\rm am}) \nonumber \\
                   &=&
8\pi (\varepsilon_{c} ) + \frac{1}{2r^3(Q^2+r^2)^2}[(-2Q^4  -3Q^2 r^2 + r^4)b(r) - r^3(Q^2+r^2) b'(r) ]  \,,  \label{Etensor3}
\end{eqnarray}
where $\varepsilon_c=\frac{Q^2}{8\pi r^4}$, $p_r$ is the radial pressure, $p_t$ is the transverse pressure,
and the prime denotes the derivative with respect to $r$.

From Eqs.\ (\ref{Etensor1}) and (\ref{Etensor2}), we obtain
\begin{equation}
b(r)=b_{o} \left(\frac{r b_{o}}{r^2+Q^2}\right)^{1/w_1}  \,.
\label{br}
\end{equation}
Substituting this result back into Eqs.\ (\ref{Etensor1}) or (\ref{Etensor2}), we get
$\varepsilon_{\rm am}=- \frac{r^2-Q^2}{8\pi w_1 r^4 }\left(\frac{b_{o} r}{r^2+Q^2}\right)^{(w_1+1)/w_1}$.
After plugging those into Eq.\ (\ref{Etensor3}), we obtain
\begin{equation}
w_2(r)=\frac{[-r^4(1+w_1)+Q^2r^2(1+3w_1)+2Q^4w_1]}{2(r^4-Q^4)} \,.
\label{eqodst2}
\end{equation}
Then, the energy density and pressures are given by
\begin{eqnarray}
\label{edenpre}
\varepsilon&=&\frac{Q^2}{8\pi r^4}-\frac{r^2-Q^2}{8\pi w_1 r^4}\left(\frac{b_{o} r}{r^2+Q^2}\right)^{(w_1+1)/w_1} \,, \nonumber \\
p_r &=& -\frac{Q^2}{8\pi r^4} -\frac{r^2-Q^2}{8\pi r^4}\left(\frac{b_{o} r}{r^2+Q^2}\right)^{(w_1+1)/w_1} \,, \\
p_t &=&  \frac{Q^2}{8\pi r^4} + \frac{1}{16\pi r^4(r^2 + Q^2)}\left(\frac{b_{o} r}{r^2+Q^2}\right)^{(w_1+1)/w_1}\frac{[r^4(1+w_1)-Q^2(r^2(1+3w_1)+2Q^2w_1)]}{w_1}\,. \nonumber
\end{eqnarray}
We could not distinctly separate the energy density and pressure contributions from the charge and the anisotropic matter.
When we set $Q=0$, the results converge to a wormhole with $Q=0$~\cite{Kim:2001ri, Halder:2019urh, Kim:2019ojs}.

\begin{figure}[H]
\begin{center}
{\includegraphics[width=4.0in]{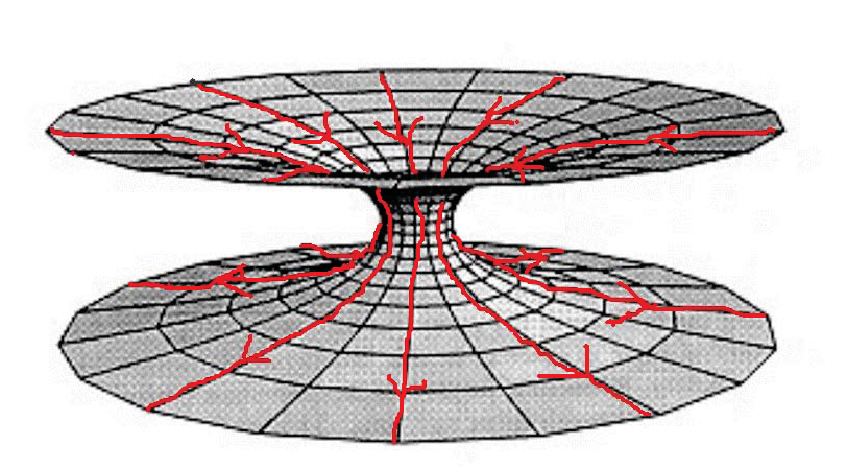}}
\end{center}
\caption{\footnotesize{(color online).
Conceptual embedded diagram of the wormhole with electric field lines.
}}
\label{wormhole}
\end{figure}
Figure \ref{wormhole} illustrates a conceptual embedded diagram of a wormhole featuring electric field lines.
In this diagram, the red electric field lines converge toward the wormhole from one universe,
traverse through it, and exit into another universe.
At the throat of the wormhole, the Maxwell tensor, $F^{tr}=\sqrt{\frac{g(r)}{f(r)}} \frac{Q}{r^2}$, goes to zero,
which ensures continuity across this region.
If one considers a Gaussian surface that surrounds the asymptotic regions of both universes,
there is an equal flux of electric field lines entering and exiting,
indicating that no net charge exists within the Gaussian surface.
Consequently, this scenario exemplifies the idea of ``charge without charge".

\subsection{Conditions for a geometry to be considered a wormhole \label{sec2-2}}

\quad

Now we outline the conditions under which the solution to the Einstein-Maxwell equation becomes a wormhole geometry.

$\bullet$ First, we need to examine the metric functions. The functions we have selected, $f(r)$ and $g(r)$,
are distinct. If they were identical~\cite{Schwarzschild:1916uq, Reissner:1916cle, Tangherlini:1963bw,
Kiselev:2002dx, Hayward:2005gi, Cho:2017nhx, Jeong:2023hom},
the radial pressure could equate to the negative energy density~\cite{Jacobson:2007tj}.
This can be readily identified by analyzing the relation $p_r+\varepsilon= \frac{g(r) f'(r)-f(r) g'(r)}{rf(r)}$.
Equations (\ref{Etensor1}) and (\ref{Etensor2}) do not satisfy this relation, except in the specific case where $w_1=-1$.

In contrast to black hole geometry, wormhole geometry does not possess an event horizon or a physical singularity.
This difference is closely tied to the properties of the functions $f(r)$ and $g(r)$.
In the case of black hole spacetime, the event horizon is defined as the point where
both $f(r)$ and $g(r)$ simultaneously go to zero.
The former indicates that the Killing vector field becomes null,
while the latter signifies that the surface at $r={\rm const}$ becomes null.
The infinite redshift surface is also obtained at the location where $f(r)$ vanishes.
There are three invariant curvature scalars to consider:
$R$, $R_{\mu\nu}R^{\mu\nu}$, and $R_{\mu\nu\alpha\beta}R^{\mu\nu\alpha\beta}$.
The denominators of these scalars are influenced by the power of $f(r)$, while $g(r)$ does not appear in the denominators.
It is feasible for $f(r)=0$ at a position where the infinite redshift surface is smaller than the wormhole throat.
Our wormhole spacetime extends from the throat location, $r=r_{o}$, to infinity.
Consequently, $f(r)$ must remain non-zero and positive for $r\geq r_{o}$.
This condition also guarantees the absence of a physical singularity for this region.

Let us examine the physical implications of spacetime at the point where the metric function $g(r)$ vanishes.
Such points occur in the geometries of both black holes and wormholes and warrant separate descriptions.
The closed two-dimensional spatial hypersurface at this location corresponds to
a marginally trapped surface for a dynamic black hole~\cite{Penrose:1964wq, Ashtekar:2004cn, Senovilla:2011fk},
whereas for a dynamic wormhole, it corresponds to a marginally anti-trapped surface~\cite{Hochberg:1998ha, Raviteja:2020fzt}.
This situation reflects a coordinate singularity.
To ensure the proper radial distance, given by $l(r)=\pm \int^r_{r_{o}} \frac{dr}{\sqrt{g(r)}}$,
remains finite everywhere, we note that for our wormhole, the condition $g(r)=0$ can yield multiple roots.
We require $g(r)$  to be non-negative near the throat and at that point.
Consequently, we select the largest root as the location of the throat,
assuming it to be the marginally anti-trapped surface.

$\bullet$ Second, we examine the flare-out condition~\cite{Morris:1988cz, Hochberg:1997wp, Hochberg:1998ha, Kim:2013tsa}
and the energy condition associated with traversable wormholes.
To construct and sustain the structure of a traversable wormhole,
the geometric flare-out condition must be satisfied at the throat and its vicinity.
This condition is intrinsically linked to the energy condition of the matter that supports the wormhole structure.
It has been noted that the divergence property of the null geodesic at the marginally anti-trapped surface
generalizes the flare-out condition~\cite{Hochberg:1998ha}.
In this paper, we analyze both the flare-out condition and the exoticity function at the throat and its vicinity,
exploring how these two conditions relate at the wormhole throat.
We begin by considering the flare-out condition of the wormhole through embedding geometry at $t=\rm const.$ and $\theta=\frac{\pi}{2}$.
This condition can be expressed as
\begin{equation}
\label{flaout2}
\frac{d^2r}{dz^2} = \frac{r[r(b(r)-rb'(r))-2Q^2]}{2(rb(r) -Q^2)^2} > 0 \,,
\end{equation}
which allows us to determine the flare-out condition by the behavior of $b(r)$ in the vicinity of the throat.

Substituting Eq.~(\ref{br}) into the flare-out condition, we get the numerator as
\begin{equation}
\label{exoticity2}
N(r) \equiv -2Q^2 +  \frac{\left( \frac{r(r^2_{o} +Q^2)}{r_{o}(r^2+Q^2)} \right)^{1+1/w_1}(Q^2(-1+w_1)+r^2(1+w_1))}{w_1} > 0 \,.
\end{equation}
At the throat, this function turns out to be
\begin{equation}
\label{exoticity22}
N(r_o) = (r^2_{o}-Q^2)\left(1+1/w_1 \right) > 0 \,.
\end{equation}
Here, we choose $w_1 > 0$ or $w_1 < -1$ for the asymptotically flat spacetime,
which requires the charge to satisfy $r_{o} > Q$ through the flare-out condition as Eq.~(\ref{exoticity22}).
We obtained the constraint for $Q$ here, and the existence of $Q$ does not modify the range of $w_1$
to satisfy the flare-out condition~\footnote{This could also be interpreted as follows: The condition for $g(r)$ to be non-negative near the throat is $g'(r_0)>0$, which is Eq.~(\ref{exoticity22}). This is automatically satisfied if the asymptotic flatness conditions $w_1 > 0$ or $w_1 < -1$ are satisfied, because $r_o$ given by Eq.~(\ref{worthroat}) is larger than $Q$ (the first term between round brackets is larger than $2Q$, and the second is positive). }.

One can introduce the exoticity function~\cite{Morris:1988cz, Kim:2013tsa} as
\begin{eqnarray}
\label{exoticity}
\zeta(r) \equiv \frac{-p_r - \varepsilon}{|\varepsilon|}
= \frac{(r^2-Q^2)(1+w_1)\left( \frac{r b_o}{r^2+Q^2} \right)^{(w_1+1)/w_1}}{8\pi r^4 w_1 |\varepsilon|}\,.
\end{eqnarray}
When the exoticity function is positive, the null energy condition is violated.
At the throat, this one becomes
\begin{eqnarray}
\label{exoticity3}
\zeta(r_o)= \frac{(r^2_o-Q^2)(1+1/w_1)}{8\pi r^4_o |\varepsilon(r_o)| } \,,
\end{eqnarray}
where $\zeta(r_o)$ takes the same form as (\ref{exoticity22}) up to a positive definite multiplication factor.
Thus, the wormhole supported by this matter could satisfy the flare-out condition.

$\bullet$ Third, let us now analyze the condition for traversability related to the tidal forces generated by the wormhole.
Near and at the wormhole, these tidal forces stretch and compress a traveler.
To ensure a safe passage, the magnitude of the tidal force experienced by the traveler must remain within tolerable limits.
We consider a traveler moving through the interior of a wormhole with a speed $u^{\hat \mu}$ in the orthonormal basis of a static observer.
Due to the inhomogeneity of the gravitational field, two points separated by the length of the traveler,
represented by the separation vector $\xi^{\hat k}$, experience differing accelerations.
This variation, denoted as $\triangle a^{\hat j}$, is what we refer to as the tidal effect.
Hence, the tidal effect can be expressed as~\cite{Morris:1988cz, Visser:1995cc, Lobo:2007zb, Shaikh:2016dpl}
\begin{eqnarray}
\triangle a^{\hat j '} = - R^{\hat j'}_{\hat 0 ' \hat k' \hat 0' } \xi^{\hat k'} \,,
\label{tidalacc}
\end{eqnarray}
where $u^{\hat \mu '}=\delta^{\hat \mu'}_{\hat 0'}$ and $\xi^{\hat 0'}=0$ in the traveler's
frame were used.
The relation between traveler's orthonormal basis $e_{\hat 0'}, e_{\hat 1'}, e_{\hat 2'}, e_{\hat 3'}$
and the static observer's orthonormal basis $e_{\hat t}, e_{\hat t}, e_{\hat \theta}, e_{\hat \phi}$
are shown in Ref.~\cite{Morris:1988cz}. The components of the Riemann tensor are given by
\begin{eqnarray}
R_{ \hat 1' \hat 0' \hat 1' \hat 0' } &=& \frac{f(r)f'(r)g'(r)-g(r)[f'^2(r)-2f(r)f''(r)]}{4f^2(r)}\,, \nonumber \\
R_{ \hat 2' \hat 0' \hat 2' \hat 0' } &=& R_{ \hat 3' \hat 0' \hat 3' \hat 0' } =  \frac{\gamma^2[g(r)f'(r)-v^2f(r)g'(r)] }{2r f(r)} \,,
\label{comreim}
\end{eqnarray}
where $\gamma=(1-v^2)^{-1/2}$ and $v=\pm \frac{\sqrt{g_{rr}}dr}{\sqrt{-g_{tt}} dt}$ is the radial velocity of the traveler.
Thus, the tidal acceleration becomes at the throat
\begin{eqnarray}
\triangle a^{\hat 1 '}|_{r_{o}} &=& \frac{Q^2(1+w_1)(r^2_{o}-Q^2)}{2w_1 r^4_{o} (r^2_{o}+Q^2)}\xi^{\hat 1'}|_{r_{o}} \,,  \nonumber \\
\triangle a^{\hat 2 '}|_{r_{o}}
&=& \triangle a^{\hat 3 '}|_{r_{o}} =  \frac{\gamma^2_o v^2_o (1+w_1)(r^2_{o}-Q^2)}{2r^4_{o} w_1} \xi^{\hat 2'}|_{r_{o}} \,,
\label{comreimatthr}
\end{eqnarray}
where $\gamma_o$ and $v_o$ denote those at the throat.
The function $f(r)$ constrains the radial component, $\triangle a^{\hat 1 '}|_{r_{o}}$,
indicating that the radial component goes to zero when $Q=0$, as expected~\cite{Morris:1988cz}.
Consequently, if the wormhole possesses a small charge $Q$,
the traveler may perceive the radial component as being within tolerable limits.
Conversely, the speed $v$ at which the traveler traverses the wormhole restricts the lateral component,
$\triangle a^{\hat 2 '}|_{r_{o}}$.
Alternatively, one could consider $|\triangle a|=\sqrt{\sum_k (\triangle a^k)^2}$ and compare it with Earth's gravity,
$g_{\bigoplus} =9.8[{\rm m/s^2}]$.
\begin{eqnarray}
\frac{|\triangle a|}{g_{\bigoplus}}
\leq |1+1/w_1|\left(1- \frac{kGQ^2}{c^4r^2_o}\right) \sqrt{\frac{k^2 G^2 Q^4}{(c^4 r^2_o +k GQ^2)^2} + 2\gamma^4_o (v_o/c)^4} \frac{c^2|\xi|}{2r^2_o g_{\bigoplus}} < 1 \,,
\label{avtiacc}
\end{eqnarray}
where we recovered the speed of light, $c=3\times 10^8 [{\rm m/s}]$, the gravitational constant, $G=6.67\times 10^{-11} [{\rm N\cdot m^2/kg^2}]$, and Coulomb's constant, $k=9 \times 10^9 [{\rm N\cdot m^2/C^2_{\rm Cou}}]$. We take the size of the traveler's body $| \xi| \sim 2 [{\rm m}]$.

\begin{figure}[h]
\begin{center}
\subfigure[The magnitude as varying $v_o$  ]
{\includegraphics[width=3.0in]{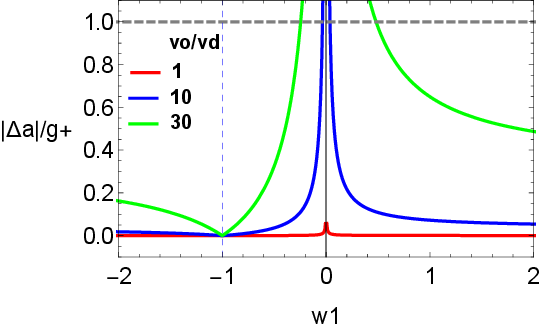}}
\subfigure[The magnitude as varying $r_o$  ]
{\includegraphics[width=3.0in]{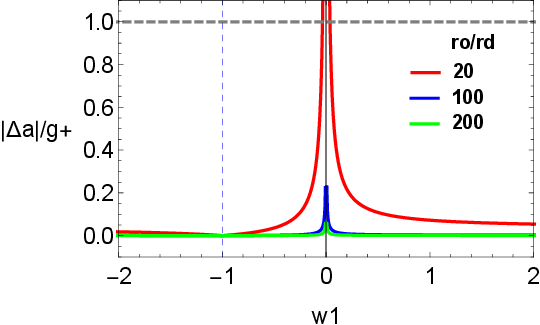}} \\
\subfigure[The magnitude as varying $Q$  ]
{\includegraphics[width=3.0in]{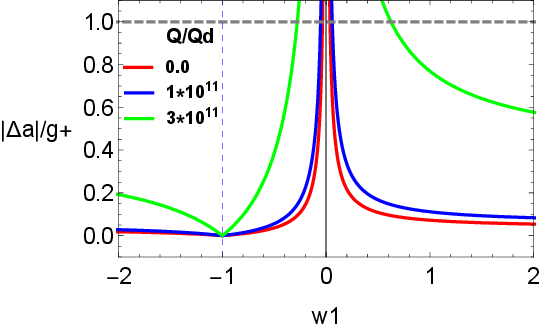}}
\subfigure[The magnitude as varying $b_o$  ]
{\includegraphics[width=3.0in]{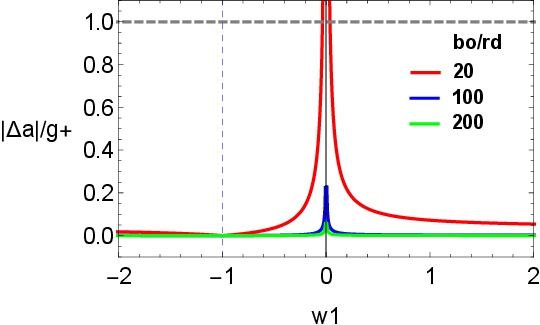}}
\end{center}
\caption{\footnotesize{(color online).
The magnitude of tidal accelerations at the throat. The $v_d$, $r_d$, and $Q_d$ are introduced to make
dimensionless quantities with the unit quantity.
}}
\label{tidalacc}
\end{figure}
Figure \ref{tidalacc} illustrates the magnitude of the tidal acceleration at the throat, as shown by Eq.~(\ref{avtiacc}).
There are four parameters that we can manipulate: the traveler's velocity $v_o$,
the size of the wormhole throat $r_o$, the charge $Q$, and the magnitude of $b_o$, respectively.
We have fixed the size of the traveler.
As indicated in Eq.~(\ref{avtiacc}), $\frac{kGQ^2}{c^4r^2_o} \ll 1$,
and the first term in the square root is significantly less than the second term.
We exclude the region where $-1\leq w_1\leq 0$, and we have marked the location of $w_1=-1$ with a dashed blue line.

In Figure \ref{tidalacc}$(a)$, we examine the magnitude of tidal acceleration for three different values of $v_o$.
The red curve corresponds to $v_o=1[{\rm m/s}]$, the blue curve to $v_o=10[{\rm m/s}]$, and the green curve to $v_o=30[{\rm m/s}]$, respectively.
We set $r_o=20 [{\rm m}]$.
As expected, the magnitude of tidal acceleration increases with the traveler's velocity.
For wormholes with $w_1<-1$, the magnitude is sufficiently small.
However, for wormholes with $w_1>0$, the magnitude can become large enough to pose a danger to the traveler at small values of $w_1$,
while remaining tolerable for larger values of $w_1$.
In Figure \ref{tidalacc}$(b)$, we examine the magnitude for three different values of $r_o$.
The red curve represents $r_o=20[{\rm m}]$, the blue curve corresponds to $r_o=100[{\rm m}]$, and the green curve to $r_o=200[{\rm m}]$, respectively.
We take $v_o=10[{\rm m/s}]$ and $Q=1[{\rm C_{ Cou}}]$.
As expected, the tidal acceleration increases as the size of the wormhole's throat decreases.
The general behavior of these curves is similar to that observed in Figure \ref{tidalacc}$(a)$.
In Figure \ref{tidalacc}$(c)$, we examine the magnitude for three different values of $Q$.
The red curve represents $Q=0[{\rm C_{ Cou}}]$, the blue curve corresponds to $Q=10^{11}[{\rm C_{ Cou}}]$,
and the green curve to $Q=3\times 10^{11}[{\rm C_{ Cou}}]$, respectively.
We maintain $v_o=10[{\rm m/s}]$ and $r_o=20 [{\rm m}]$.
As expected, the tidal acceleration increases with increasing $Q$.
If we continue to increase $Q$ such that the decrease in the term in front of the square root is significant relative to
the increase in the square root in Eq.~(\ref{avtiacc}),
the tidal acceleration could decrease again.
However, we did not examine this possibility because $Q$ becomes excessively large.
In Figure \ref{tidalacc}$(d)$, we examine the magnitude for three different values of $b_o$.
The red curve corresponds to $b_o=20[{\rm m}]$, the blue curve to $b_o=100[{\rm m}]$,
and the green curve to $b_o=200[{\rm m}]$, respectively.
We keep $v_o=10[{\rm m/s}]$ and $Q=1[{\rm C_{ Cou}}]$.
The curves exhibit almost the same behavior.

\subsection{Deflection angle, photon sphere, and capture cross-section \label{sec2-3}}

\quad  We now examine the geodesics of light around the wormhole geometry,
focusing on how light is deflected by the wormhole and
how this affects the photon sphere and the capture cross-section in comparison to that of a black hole.

The geometry of a static spherically symmetric wormhole is anticipated to exhibit
the same spacetime symmetry as that of a static spherically symmetric black hole.
Specifically, it is expected to possess four Killing vectors and two Killing tensors within four-dimensional spacetime.
Among these, two Killing vectors and two Killing tensors in the Hamilton-Jacobi formalism eventually provide four first-order differential equations
that correspond to the geodesic equations~\cite{Walker:1970un, Misner:1973prb, Benenti:1979erw, Frolov:2017kze, Bezdekova:2022gib, Jeong:2023hom}.
Given the spherically symmetric nature of the geometry, we can confine our analysis to
equatorial orbits by setting $\theta=\pi/2$, without any loss of generality.
They are given by
\begin{eqnarray}
&& \frac{dt}{d\lambda}  = \frac{E}{f(r)}    \,, \label{HJeq-t}\\
&& \frac{d\phi}{d\lambda} =  \frac{L_z}{r^2}   \,, \label{HJeq-phi} \\
&& \frac{dr}{d\lambda} = \pm \sqrt{g(r)\left[ \frac{E^2}{f(r)} - \frac{L^2_z}{r^2} \right] }  \,,\label{HJeq-the}
\end{eqnarray}
where $E$ and $L_z$  correspond to the energy and the angular momentum of the light at infinity.
The signs $+1(-1)$ in Eq.~(\ref{HJeq-the}) corresponds to the outgoing (ingoing) geodesics.
From Eq.~(\ref{HJeq-the}), we obtain  the radial equation for    the geodesic motion
\begin{equation}
\label{raged}
\frac{1}{2}\left(\frac{dr}{d\lambda} \right)^2  +   V_{\rm eff} =0 \,,
\end{equation}
where $V_{\rm eff} = \frac{g(r)}{2}\left[ - \frac{E^2}{f(r)} + \frac{L^2_z}{r^2} \right] $.
We now examine where a photon sphere exists.
Like black holes, wormholes also possess a gravitational field that could bend the path of light,
causing it to orbit the wormhole.
This phenomenon is referred to as a photon sphere.
The photon sphere is determined by the radii of
unstable circular orbits, i.e., $V_{\rm eff}|_{r_p}=0$, $\frac{d V_{\rm eff}}{ dr}|_{r_p}=0$, and $\frac{d^2 V_{\rm eff}}{ dr^2}|_{r_p} < 0$.
Since $g(r)=0$ at the position $V_{\rm eff}=0$, both conditions are fulfilled
if the square bracket quantity is also zero at this position.
Note that the photon sphere exists at the wormhole throat.

\begin{figure}[h]
\begin{center}
\subfigure[The effective potential as varying $L_z$  ]
{\includegraphics[width=3.0in]{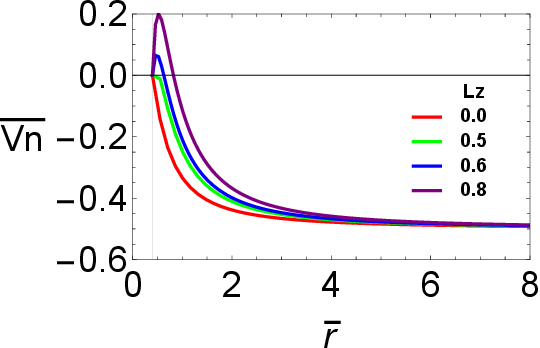}}
\subfigure[The effective potential for the photon sphere as varying $w$  ]
{\includegraphics[width=3.0in]{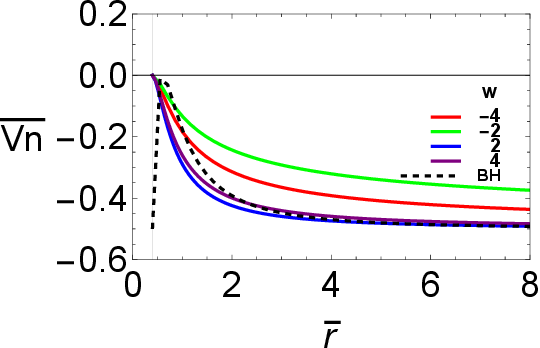}}
\end{center}
\caption{\footnotesize{(color online) The effective potential
}}
\label{phospe}
\end{figure}
Figure \ref{phospe} represents the shape of the effective potential as defined by Eq.~(\ref{raged}).
We take $Q/r_d=0.2$, $b_o/r_d=0.5$, and $E=1$, in which $r_d$ is introduced for making the dimensionless quantities.
Figure \ref{phospe}~(a) illustrates how the shape of the effective potential changes with variations in the angular momentum $L_z$.
We take $w=2$, the red curve with $L_z/r_d=0$, the green curve with $L_z/r_d=0.5$, the blue curve with $L_z/r_d=0.6$,
and the purple curve with $L_z/r_d=0.8$, respectively.
The photon sphere is formed at specific values of $L_z$,
where the local maximum of the effective potential is zero.
For all other values of $L_z$, there are no stable orbits that rotate the wormhole.
Instead, light that approaches the wormhole is bent by its gravitational field and will move away from it.
Figure \ref{phospe}~(b) illustrates how the shape of the effective potential changes with variations of $w$,
with $L^2_z = \frac{E^2 r^3_o}{b_o}$ adjusted to ensure that the local maximum of the effective potential is set to zero.
For reference, the case of a Schwarzschild black hole is represented by black dotted lines,
where the location of the black hole horizon $r_H$ aligns with that of the wormhole throat.
In the case of a Schwarzschild black hole with,
the local maximum of the effective potential is obtained at $r_{\rm pb}=\frac{3}{2} r_H$.
Notably, in a black hole geometry, the photon sphere is positioned outside the horizon,
while for a wormhole, it is situated at the throat.
We take the red curve with $w=-4$, the green curve with $w=-2$, the blue curve with $w=2$, and the purple curve with $w=4$, respectively.
The overall shape of the effective potential remains relatively similar across different values of $w$.

We now determine the maximum radius at which light coming from infinity to the wormhole will be captured eventually.
From Eqs.~(\ref{HJeq-phi}) and (\ref{HJeq-the}), we get
\begin{equation}
\label{drdphi}
\frac{dr}{d\phi} = \pm \sqrt{g(r)\left[ \frac{E^2 r^4}{f(r) L^2_z} - r^2\right] }  \,.
\end{equation}
To facilitate this, we introduce the impact parameter $r_{\rm ip}$,
which defines the area of the capture cross-section as $\sigma_{\rm ip}=\pi r^2_{\rm ip}$~\cite{Dymnikova1986}.
The impact parameter is determined as the ratio of the angular momentum of the light to its energy at infinity,
such that $L_z = E r_{\rm ip}$.
This relationship is equivalent to the condition where Eq.~(\ref{drdphi}) yields zero when $r_{\rm ip} = L_z/E$ is substituted into Eq.~(\ref{drdphi}).
Consequently, we obtain that $r_{\rm ip} = \sqrt{\frac{r^3_o}{b_o}}$, regardless of other parameter values.
For a Schwarzschild black hole, the impact parameter is given by $r_{\rm ipb}=\frac{3\sqrt{3}}{2} r_H $.
When light comes from infinity, the impact parameter corresponds to the boundary that determines
whether the light will be captured. This distance is therefore also referred to
as the radius of the black hole shadow~\cite{EventHorizonTelescope:2020qrl}.
\begin{figure}[h]
\begin{center}
{\includegraphics[width=3.5in]{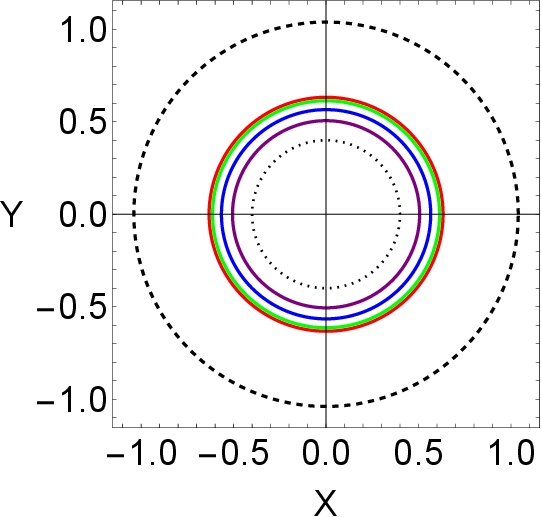}}
\end{center}
\caption{\footnotesize{(color online) The capture cross-section of light by a wormhole.
}}
\label{capture}
\end{figure}
Figure \ref{capture} represents the capture cross-section of light by a wormhole.
The positions of the wormhole throat and the event horizon are set to $r_o/r_d=r_H/r_d=0.4$, represented by a black dotted curve.
Using a Schwarzschild black hole as a reference, the impact parameter is indicated by the black dashed curve.
For the wormhole, the red curve corresponds to $Q/r_d = 0$ and $r_{\rm ip}/r_d \simeq 0.6325$,
the green curve to $Q/r_d = 0.1$ and $r_{\rm ip}/r_d \simeq 0.6136$,
the blue curve to $Q/r_d = 0.2$ and $r_{\rm ip}/r_d \simeq 0.5659$,
and the purple curve to $Q/r_d = 0.3$ and $r_{\rm ip}/r_d \simeq 0.5060$.
Among these, the Schwarzschild black hole has the largest capture cross-section,
whereas the wormhole has a larger capture cross-section for smaller values of $Q$ and larger values of $b_o$ for fixing $r_o$.

Research has been conducted into finding the deflection angle in the gravitational field of a wormhole~\cite{Tsukamoto:2012xs, Shaikh:2018oul},
similar to that of a black hole~\cite{Weinberg:1972kfs, Virbhadra:1999nm, Bozza:2002zj}.
Light that comes from infinity reaches a radial turning point, and then usually returns to infinity.
Our aim now is to determine how the geometry of our wormhole solution affects the deflection angle comparing to that by a black hole.
It is given by~\cite{Weinberg:1972kfs}
\begin{eqnarray}
\label{deflecan}
\alpha = 2 \int^{\infty}_{r_{p}} \frac{\sqrt{f(r)}}{r^2 \sqrt{g(r)} \sqrt{\frac{f(r_{b})}{r^2_{b}} - \frac{f(r)}{r^2} } } - \pi\,,
\end{eqnarray}
where $r_{b}$ corresponds to the distance that is closest to the object.
We adjust $r_{b}$ to identify the point at which the deflection angle diverges.
\begin{figure}[h]
\label{deflection}
\begin{center}
{\includegraphics[width=3.5in]{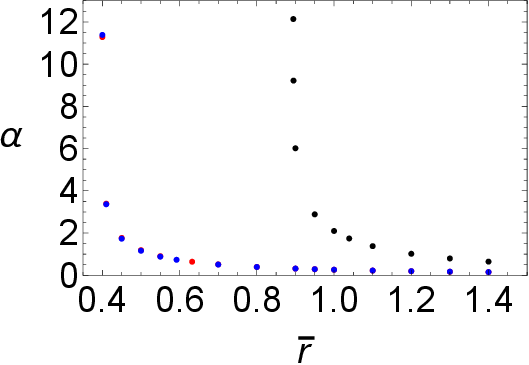}}
\end{center}
\caption{\footnotesize{(color online) The deflection angle by a wormhole.
}}
\end{figure}
Figure \ref{capture} represents the deflection angle by a wormhole.
We take $r_o/r_d=0.4$ and $w=2$, the red dot with $Q/r_d=0$, and the blue dot $Q/r_d=1.5$.
These two points do not greatly affect the appearance of the graph.
As the $r_b$ value approaches the position of the photon sphere, i.e., at the throat,
the deflection angle shows a tendency to diverge in the wormhole geometry~\cite{Shaikh:2018oul},
while it shows a tendency to diverge before reaching the photon sphere in the Schwarzschild black hole geometry.

\section{Trying to construct charged rotating wormholes \label{sec3}}
\quad

In this section, we construct charged rotating wormhole geometries from the static solutions via the modified NJ algorithm.
It may be useful to compare this section with the rotating black hole case described in Appendix~[\ref{app:robh}].

\subsection{The modified NJ algorithm \label{sec3-1}}

\quad

We begin with the retarded Eddington-Finkelstein coordinates
\begin{eqnarray}
ds^2&=& - f(r) du^2 -\frac{2\sqrt{f(r)} dudr}{\sqrt{g(r)}} + r^2 d\Omega^2 \,.
\label{edfin}
\end{eqnarray}
We take the null tetrad $\{ l_{\mu}, n_{\mu}, m_{\mu}, \bar{m}_{\mu}  \}$ consisted of two real null vectors,
$l_{\mu}$ and $n_{\mu}$, and a pair of complex null vectors, $ m_{\mu}$ and $\bar{m}_{\mu}$. The tetrad satisfies
the pseudo-orthogonality relations $l^{\mu}n_{\mu}=-1$, $m^{\mu}\bar{m}_{\mu}=1$, and
$l^{\mu}m_{\mu} = l^{\mu}\bar{m}_{\mu} = n^{\mu}m_{\mu} = n^{\mu}\bar{m}_{\mu} = l^{\mu}l_{\mu}
 = n^{\mu}n_{\mu} = m^{\mu}m_{\mu} =\bar{m}^{\mu} \bar{m}_{\mu} = 0$.
The metric has the relation with the null tetrad as
\begin{equation}
g_{\mu\nu}= -l_{\mu}n_{\nu} -n_{\mu}l_{\nu} + m_{\mu}\bar{m}_{\nu} +\bar{m}_{\mu}m_{\nu} \,. \label{metricstandecom}
\end{equation}
We can read off the component of the tetrad from the metric (\ref{edfin})
\begin{eqnarray}
l_{\mu}&=&  \delta^0_{\mu}\,,~~ n_{\mu} = \frac{f(r)}{2}\delta^0_{\mu} + \sqrt{\frac{f(r)}{g(r)}} \delta^1_{\mu}\,, \nonumber \\
m_{\mu}&=& \frac{r}{\sqrt{2}} \left(\delta^2_{\mu} + i\sin\theta \delta^3_{\mu} \right) \,,~~
\bar{m}_{\mu}= \frac{r}{\sqrt{2}} \left(\delta^2_{\mu} - i\sin\theta \delta^3_{\mu}\right)  \,, \label{tetrad00}
\end{eqnarray}
and
\begin{eqnarray}
l^{\mu}&=& -\sqrt{\frac{g(r)}{f(r)}}\delta^{\mu}_1 \,,~~ n^{\mu}= -\delta^{\mu}_0 +\frac{\sqrt{f(r)g(r)}}{2}\delta^{\mu}_1\,,   \nonumber \\
m^{\mu}&=&  \frac{1}{\sqrt{2}r} (\delta^{\mu}_2 + \frac{i}{\sin\theta} \delta^{\mu}_3) \,,~~
\bar{m}^{\mu}= \frac{1}{\sqrt{2}r} (\delta^{\mu}_2 -\frac{i}{\sin\theta} \delta^{\mu}_3)  \,. \label{tetrad01}
\end{eqnarray}

We now perform the transformations
\begin{equation}
u'= u- ia\cos\theta,~~~ r' = r +ia\cos\theta \,,
\label{cortt}
\end{equation}
where $a$ is a rotation parameter.
We take the below for a rotating wormhole, unlike a rotating black hole
\begin{eqnarray}
\label{whrho}
r^2  \Rightarrow {\bar\rho}^2 = r'^2 - a^2\cos^2\theta  \,,~~f(r) \Rightarrow F(r', \theta)\,,~~g(r) \Rightarrow G(r', \theta) \,.
\end{eqnarray}
For the rotating black hole, ${\bar\rho}^2_{\rm BH} = r'^2 + a^2\cos^2\theta $.
For comparison with the geometry of a rotating wormhole, the geometry of a rotating black hole with the proper
NJ algorithm is described in the Appendix.

Here after, we omit the prime and the functions $F(t, \theta)$ and $G(r, \theta)$ are given by
\begin{equation}
\label{metro}
F(t, \theta) = 1+ \frac{Q^2}{{\bar\rho}^2} \,,~~~ G(r, \theta) = 1+ \frac{Q^2}{{\bar\rho}^2}
- \frac{b_{a}r}{{\bar\rho}^2} \left( \frac{b_{a} r}{r^2 +Q^2}  \right)^{\frac{1}{w_1}} \,,
\end{equation}
where $b_{a}$ will be specified later.
And then the tetrad becomes
\begin{eqnarray}
\label{tetrad3}
l^{\mu}&=& -\sqrt{\frac{G(r, \theta)}{F(r, \theta)}}\delta^{\mu}_1 \,, ~~
n^{\mu}=-\delta^{\mu}_0 +\frac{\sqrt{F(r, \theta)G(r, \theta)}}{2}\delta^{\mu}_1   \nonumber \\
m^{\mu}&=&\frac{\left(ia\sin\theta(\delta^{\mu}_0 - \delta^{\mu}_1)+\delta^{\mu}_2 +\frac{i}{\sin\theta}\delta^{\mu}_3\right)}{\sqrt{2 \bar\rho}}  \,, ~~
\bar{m}^{\mu}= \frac{\left(-ia\sin\theta(\delta^{\mu}_0-\delta^{\mu}_1)+\delta^{\mu}_2 -\frac{i}{\sin\theta}\delta^{\mu}_3\right)}{\sqrt{2 \bar\rho}}  \,.
\end{eqnarray}
From the tetrad in (\ref{tetrad3})
\begin{displaymath}
\label{inverseme}
g^{\mu\nu} = \left(\begin{array}{cccc}
\frac{a^2\sin^2\theta}{r^2- a^2\cos^2\theta} & - \sqrt{\frac{G(r, \theta)}{F(r, \theta)}} - \frac{a^2\sin^2\theta}{r^2- a^2\cos^2\theta} & 0  &  \frac{a}{r^2- a^2\cos^2\theta} \\
- \sqrt{\frac{G(r, \theta)}{F(r, \theta)}} - \frac{a^2\sin^2\theta}{r^2- a^2\cos^2\theta} &  G(r, \theta)  + \frac{a^2\sin^2\theta}{r^2- a^2\cos^2\theta} & 0 & - \frac{a}{r^2- a^2\cos^2\theta} \\
0  & 0 & \frac{1}{r^2- a^2\cos^2\theta} & 0 \\
\frac{a}{r^2- a^2\cos^2\theta} & - \frac{a}{r^2- a^2\cos^2\theta} & 0 & \frac{1}{(r^2- a^2\cos^2\theta)\sin^2\theta}
\end{array}  \right)\,,
\end{displaymath}

\subsection{Useful coordinates \label{sec3-2}}

\quad
The Eddington-Finkelstein form of the geometry is
\begin{eqnarray}
ds^2=&& - F(r, \theta) du^2 -2\sqrt{\frac{F(r, \theta)}{G(r, \theta)}} dudr + 2 \sqrt{\frac{F(r, \theta)}{G(r, \theta)}} a\sin^2\theta dr d\psi
 \nonumber \\
&& -2 \left(\sqrt{\frac{F(r, \theta)}{G(r, \theta)}} -F(r, \theta)\right) a\sin^2\theta dud\psi + {\bar\rho}^2 d\theta^2 + \Sigma_{WH} \sin^2\theta d\psi^2 \,,
\end{eqnarray}
where
\begin{eqnarray}
\Sigma_{WH}  = \left({\bar\rho}^2 +a^2 \sin^2\theta \left(2\sqrt{\frac{F(r, \theta)}{G(r, \theta)}} - F(r, \theta) \right)\right) \,.
\end{eqnarray}
Then the null the tetrad becomes
\begin{eqnarray}
\label{tetrad4}
l_{\mu}&=& \delta^0_{\mu} - a\sin^2\theta \delta^3_{\mu}   \,, ~~
n_{\mu}=\frac{F(r, \theta)}{2}\delta^0_{\mu} + \sqrt{\frac{F(r, \theta)}{G(r, \theta)}}\delta^1_{\mu} + a\sin^2\theta \left( \sqrt{\frac{F(r, \theta)}{G(r, \theta)}} - F(r, \theta)\right) \delta^3_{\mu}   \nonumber \\
m_{\mu}&=&\frac{\bar\rho}{\sqrt{2} }\left[\delta^2_{\mu} + i\sin\theta \delta^3_{\mu} \right]   \,, ~~
\bar{m}_{\mu}= \frac{\bar\rho}{\sqrt{2} }\left[\delta^2_{\mu} - i\sin\theta \delta^3_{\mu} \right] \,.
\end{eqnarray}

We try to use
\begin{eqnarray}
\label{difwh1}
du=  dt - \frac{A(r, \theta)}{\triangle_2} dr + B(r, \theta) d \phi\,, ~~~
d\psi= d\phi - \frac{a[1+C(r, \theta)]}{\triangle_2} dr\,,
\end{eqnarray}
to obtain Boyer-Lindquist coordinates, where
\begin{eqnarray}
\label{difwh02}
\vartriangle_2  ={\bar\rho}^2 G(r,\theta) - a^2 \sin^2\theta \,.
\end{eqnarray}
For the rotating black hole, $\vartriangle_{\rm BH 2} = {\bar\rho}^2_{\rm BH} G(r,\theta) + a^2 \sin^2\theta $.
We restrict the metric function to have $dtdr=0$ and $drd\phi=0$.
Under these restrictions, we find the forms of the metric functions $A(r, \theta)$,
$B(r, \theta)$, and $C(r, \theta)$ as follows:
\begin{eqnarray}
\label{metricfunc02}
A(r, \theta) &=& - \frac{\vartriangle_2}{\sqrt{F(r, \theta)G(r, \theta)}}- a^2\sin^2\theta(1+C(r, \theta))\left(1- \frac{1}{\sqrt{F(r, \theta)G(r, \theta)}}\right) \,, \nonumber \\
B(r, \theta) &=& -\frac{2a\sin^2\theta [C(r, \theta)\vartriangle_2 + 2a^2\sin^2\theta (1+C(r, \theta)) ] }{G(r,\theta)\left[\sqrt{\frac{F(r, \theta)}{G(r, \theta)}}\vartriangle_2 + a^2\sin^2\theta(1+C(r, \theta)) \left( F(r, \theta) - \sqrt{\frac{F(r, \theta)}{G(r, \theta)}} \right)   \right] } \,,  \\
C(r, \theta) &=& - \frac{2a^2\sin^2\theta\pm \sqrt{2}\sqrt{{\bar\rho}^2 G(r,\theta)\vartriangle_2 }}{{\bar\rho}^2 G(r,\theta)+ a^2 \sin^2\theta} \,. \nonumber
\end{eqnarray}
Since we do not know the metric function of a rotating wormhole that we could use as a reference,
such as a Kerr black hole for the rotating black hole, let us try with this.
This choice of Eq.~(\ref{difwh1}) gives a simple form to the $g_{rr}$ function in Eq.~(\ref{fullmetric02})
and provides reasonable results when investigating the flare-out condition.
One can also see the metric function for a rotating wormhole using a prescription used in Ref.~\cite{Azreg-Ainou:2015pda, Azreg-Ainou:2011tkx}.
Hopefully, these various attempts could help one find the rotating wormhole solution.

Then, the Boyer-Lindquist coordinates is given by
\begin{eqnarray}
\label{fullmetric02}
ds^2 &=& - F(r, \theta) dt^2 - 2\left[a\sin^2\theta \left( \sqrt{\frac{F(r, \theta)}{G(r, \theta)}} -F(r, \theta) \right) + F(r, \theta)B(r, \theta) \right] dt d\phi \nonumber \\
 && + \frac{{\bar\rho}^2}{\vartriangle_2} dr^2   + {\bar\rho}^2 d\theta^2  + D(r, \theta) d\phi^2  \label{fullmetric} \\
 &=& - \frac{\triangle_3}{{\bar\rho}^2} (dt -a \sin^2\theta \triangle_4 d\phi)^2 + \frac{\sin^2\theta}{{\bar\rho}^2}[(r^2+a^2)\triangle_5 d\phi -a dt]^2 \nonumber \\
&& + \frac{{\bar\rho}^2}{\vartriangle_2} dr^2   + {\bar\rho}^2 d\theta^2 \,,
\end{eqnarray}
where
\begin{eqnarray}
\label{metricfunc04}
&&D(r, \theta) = \left[ \Sigma_{WH} \sin^2\theta +B(r, \theta) \left[F(r, \theta)B(r, \theta)- 2a\sin^2\theta\left( \sqrt{\frac{F(r, \theta)}{G(r, \theta)}} -F(r, \theta)\right)  \right] \right]  \,,  \nonumber \\
&&\vartriangle_3 =  {\bar\rho}^2 F(r,\theta) + a^2 \sin^2\theta \,, \nonumber \\
&&\vartriangle_4 = 1 - \sqrt{\frac{1}{F(r,\theta)G(r,\theta)}} - \frac{B(r, \theta)}{a\sin^2\theta}
\pm \sqrt{\frac{{\bar\rho}^2G(r, \theta)+a^2\sin^2\theta}{\vartriangle_3 F(r, \theta)G(r, \theta)}}\,, \nonumber \\
&&\vartriangle_5 = - \frac{a B(r, \theta)}{(r^2+a^2) } + \frac{a^2\sin^2\theta\left(1-\sqrt{\frac{1}{F(r, \theta)G(r, \theta)}} \right)}{(r^2+a^2)} \pm \frac{1}{(r^2+a^2)} \sqrt{\frac{\vartriangle_3({\bar\rho}^2G(r, \theta)+a^2\sin^2\theta)}{ F(r, \theta)G(r, \theta)}}  \,.
\end{eqnarray}
When vanishing $a$, Eq.~(\ref{fullmetric}) reduces to Eq.~(\ref{stmetric}) with Eq.~(\ref{difwh1}).

We attempted to obtain the Einstein equations using the Mathematica program,
but the results of our calculations were so voluminous that we felt it would not be useful to write them all down.
To confirm that the geometry of the obtained wormhole is due to a rotating object,
the $G_{r\theta}$ component must be zero~\cite{Azreg-Ainou:2014aqa}.
However, we could not demonstrate that this is the case for our wormhole.

The determinant factor is
\begin{eqnarray}
\label{deterfacwh}
{\rm SDet_{ro}}= {\bar\rho}^2 \sqrt{\frac{F(r,\theta)(2F(r,\theta)G(r,\theta)B(r,\theta)^2+\sin^2\theta({\bar\rho}^2G(r,\theta)+a^2\sin^2\theta))}{G(r,\theta)\vartriangle_2}} \,.
\end{eqnarray}
When vanishing $a$, it turns out to be ${\rm SDet_{nro}}=r^2\sin\theta \sqrt{\frac{f(r)}{g(r)}}$, as expected.

From Eq.~(\ref{fullmetric}), the set of covariant tetrad(co-tetrad) is as follows:
\begin{eqnarray}
\label{cotetrad}
&&e^{\hat t}_{\mu} = \frac{\sqrt{\vartriangle_3}}{{\bar\rho}} (1,0,0,-a\sin^2\theta \vartriangle_4)\,, ~~~
e^{\hat r}_{\mu} = \frac{{\bar\rho}}{\sqrt{\vartriangle_2}} (0,1,0,0) \,, \nonumber \\
&& e^{\hat \theta}_{\mu} = {\bar\rho} (0,0,1,0) \,, ~~~
e^{\hat \phi}_{\mu} =  \frac{\sin\theta}{{\bar\rho}} (-a,0,0, (r^2+a^2)\vartriangle_5)  \,.
\end{eqnarray}

\subsection{Conditions for a geometry to be considered a wormhole \label{sec3-3}}

\quad

We now describe the conditions for the solution, Eqs.~(\ref{metro}) and (\ref{fullmetric}), to be considered a wormhole geometry.

$\bullet$ First, let us examine the metric functions. The metric function, $F(r, \theta)$, is non-zero and positive
everywhere, indicating that this geometry does not possess an infinite redshift surface.
In the Kerr black hole geometry, $F(r, \theta)=G(r, \theta)$, and the location defined by $G(r, \theta)=0$ corresponds to that of the ergosphere,
while the location determined by  $\vartriangle_2 =0$ indicates that of the event horizon.
Consequently, the location given by $G(r, \theta)=0$ is situated at a greater radius than that given by $\vartriangle_2 =0$;
the two locations only coincide at $\theta=0$ and $\theta=\pi$.
Importantly, there are no physical singularities outside the event horizon.
In applying these guidelines to a wormhole,
we observe that the Boyer-Lindquist coordinates provided by Eq.~(\ref{fullmetric})
lead to a divergence in the metric function terms, $\sqrt{\frac{F(r, \theta)}{G(r, \theta)}}$, before reaching the wormhole throat.
This divergence results in a physical singularity occurring outside the throat.
To circumvent this issue, Eqs.~(\ref{whrho}), (\ref{difwh1}), and (\ref{difwh02}) must differ from their black hole counterparts.
For the rotating wormhole, the location where $\vartriangle_2$ approaches zero
is located further out than where $G(r,\theta)$ approaches zero.
Nevertheless, these equations are identical at $\theta=0$ and $\theta=\pi$,
which gives rise to physical singularities at the northern and southern poles.
We will discuss this in the final section.

The $b_a$ is specified at the throat:
\begin{equation}
b_{a}=\frac{1}{r_o} \left( r^2_{o} -a^2 +Q^2 \right)^{\frac{w_1}{1+w_1}}\left( r^2_{o}+Q^2\right)^{1/(1+w_1)} \,,
\end{equation}
where $b_a$, $Q$, and $a$ are parameters describing the rotating wormhole geometry,
and their quantities determine the size of the wormhole throat.
The rotation effect serves to increase the size of the throat,
whereas the charge effect tends to decrease it. Due to its complexity,
it cannot be represented as simply as the defining equation for $r_o$.
When $a$ is set to zero, the equation simplifies to Eq.~(\ref{worthroat}).
Notably, the throat's location is independent of any angular dependence,
as referenced in Refs.~\cite{Teo:1998dp, Lobo:2007zb}.

$\bullet$ Second, let us analyze the flare-out condition at $t=\rm const.$ and $\theta=\frac{\pi}{2}$:
\begin{eqnarray}
ds^2_{\rm eq} =
\left[ 1+ \left( \frac{dz}{dr} \right)^2   \right] dr^2 + r^2_a d\phi^2 \,,
\label{2metric}
\end{eqnarray}
where $r^2_a= r^2 +a^2 \left( 2\sqrt{\frac{f(r)}{g(r) }} -f(r)  \right) + B(r)\left[f(r)(B(r)+2a)-2a\sqrt{\frac{f(r)}{g(r)}}\right]$
with $B(r)$ is from $B(r,\theta)$ with $\theta=\frac{\pi}{2}$.

The condition is given by the minimality of the throat as
\begin{equation}
\frac{d^2r}{dz^2}   > 0 \,.
\label{flareoutrot00}
\end{equation}
In Eq.~(\ref{flareoutrot00}), the denominator is always greater than zero,
thus we only examine the contribution of the numerator.
\begin{equation}
FA(w_1) \equiv (r^2_o+Q^2)(r^2_o-Q^2)(1+1/w) + a^2[r^2_o(1-1/w)+Q^2(1+1/w)] >0 \,,
\label{flareoutrot1}
\end{equation}
where we set $FA(w_1)$ to be the function of checking the flare-out condition.
When $a$ vanishes, it reduces to Eq.~(\ref{exoticity22}), while $Q$ vanishes, it reduces to
\begin{equation}
r^2_o\frac{(w_1+1)}{w_1} + a^2\frac{(w_1 -1)}{w_1} > 0 \,.
\end{equation}
If we take $(r^2_o-a^2) > 0$, then $w_1 >0$ or $w_1 < -1 +\frac{2a^2}{r^2_o+a^2}$.
For the non-vanishing $a$ and $Q$, we take $r^2_o-Q^2>0$ and $r^2_o -a^2 >0$
then the flare-out condition is satisfied with the range
\begin{equation}
w_1 >0 ~~{\rm or}~~w_1 < -1 +\frac{2a^2 r^2_o}{(r^2_o+a^2-Q^2)(r^2_o+Q^2)}\,.
\label{foarpw2}
\end{equation}
We conduct numerical calculations to investigate how the rotation of a wormhole
influences the flare out condition and to observe those behaviors.

\begin{figure}[H]
\begin{center}
\subfigure[With varying $r_o$  ]
{\includegraphics[width=3.in]{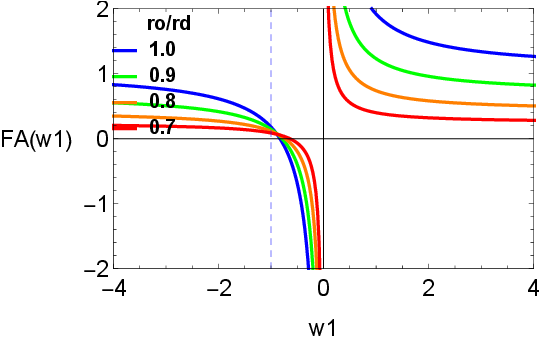}}
\subfigure[With varying $a$  ]
{\includegraphics[width=3.in]{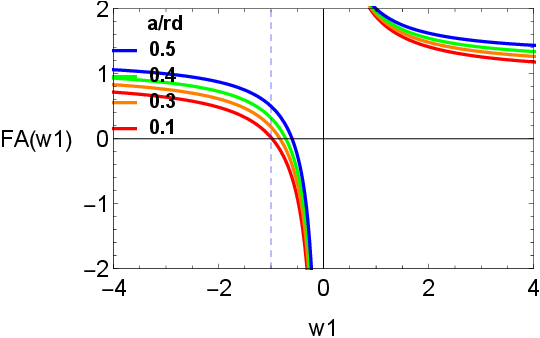}}
\subfigure[With varying $Q$  ]
{\includegraphics[width=3.in]{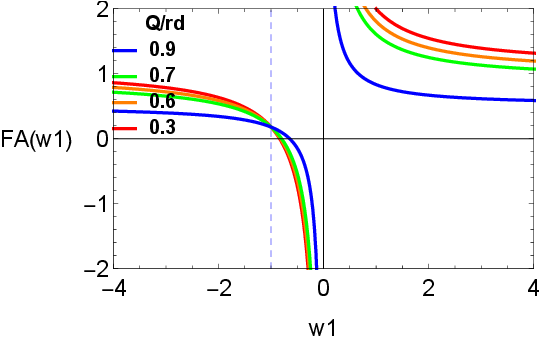}}
\end{center}
\caption{\footnotesize{(color online).
Plots to show the flare-out condition satisfied by varying the parameters $r_o$, $a$, and $Q$ for the rotating wormhole.
}}
\label{flareoutrot}
\end{figure}
Figure \ref{flareoutrot} illustrates several specific cases of the flare-out condition for the rotating wormhole.
The vertical axis corresponds to Eq.~(\ref{flareoutrot1}),
while the horizontal axis represents $w_1$.
The dashed blue line indicates the location of $w_1=-1$.
We have plotted four different values for $r_o/r_d$, $a/r_d$, and $Q/r_d$,
using red, orange, green, and blue lines, respectively.
In Figure \ref{flareoutrot}(a), we examine the flare-out condition with varying $r_o$, taking $a/r_d= 0.3$ and $Q/r_d=0.5$.
Here, the blue curve represents $r_o/r_d =1.0$, the green curve corresponds to $r_o/r_d =0.9$,
the orange curve indicates $r_o/r_d =0.8$, and the red curve signifies $r_o/r_d =0.7$.
Figure \ref{flareoutrot}(b) depicts the flare-out condition with varying $a$,
where we set $r_o/r_d=1$ and $Q/r_d=0.5$. In this case, the blue curve represents $a/r_d=0.5$,
the green curve indicates $a/r_d=0.4$, the orange curve corresponds to $a/r_d=0.3$, and the red curve signifies $a/r_d=0.1$.
Figure \ref{flareoutrot}(c) illustrates the flare-out condition with varying $Q$,
with $r_o/r_d=1$ and $a/r_d= 0.3$.
The blue curve represents $Q/r_d=0.9$, the green curve indicates $Q/r_d=0.7$,
the orange curve corresponds to $Q/r_d=0.6$, and the red curve signifies $Q/r_d=0.3$.
Since the wormhole geometry that satisfies the flare-out condition necessitates that $FA(w_1)$ be greater than zero,
we derive a constraint on the parameter values of the equation of state as expressed in Eq.~(\ref{foarpw2}).

\section{ Summary and discussions \label{sec4}}
\quad

We presented new charged wormhole solutions that is supported by anisotropic matter fields
within an asymptotically flat spacetime.
To achieve this, we employed a static, spherically symmetric ansatz with a charge
and derived the solution from the Einstein-Maxwell equations.
The anisotropic matter is characterized by two equations of state parameters:
one, denoted as $w_1$, is a constant, while the other, $w_2$, is expressed as a function of $r$,
along with the constant $w_1$ and the charge $Q$.

We then analyzed the conditions necessary for our solutions to represent a traversable wormhole geometry.
During our examination of the metric functions,
we made several observations regarding how this geometry differs from that of a black hole.
First of all, the wormhole geometry lacks an event horizon and does not exhibit singularities within its structure.
We discussed the physical implications of the points at which the metric functions $f(r)$ and $g(r)$
become zero.
For the wormhole geometry, when considering an asymptotically flat geometry,
we determined that the magnitude of the charge $Q$ must be less than the size of the wormhole throat.
The condition imposed on the asymptotically flat geometry for $w_1$ is the same as the condition for $w_1$,
which gives the value of the wormhole geometry that satisfies the flare-out condition.
We also examined the exoticity function, which is directly related to the violation of the null energy condition.
We also analyzed the magnitudes of tidal accelerations at the wormhole throat in the figure \ref{tidalacc},
in which we recovered the speed of light $c$, the gravitational constant $G$, and Coulomb's constant $k$ in SI units.
We showed that the tidal effects is small enough so that a traveler through the wormhole would have stable travel.

We analyzed the geodesics of light in the wormhole geometry that we obtained.
In particular, we examined the effective potential and investigated the possibility of unstable circular light orbits.
Our results showed that a photon sphere could exist at the location of the wormhole throat.
Additionally, we examined the position of the capture cross-section of light by the wormhole.
While we cannot definitively prove the existence of astrophysical wormholes in our Universe at this time,
these studies illustrate how wormholes can manifest phenomena that differ from those associated with black holes.

Numerous objects in the Universe exhibit rotation,
making the construction of a rotating wormhole geometry particularly intriguing.
Teo was the first to derive a rotating wormhole geometry~\cite{Teo:1998dp},
which has since sparked a plethora of studies on this concept~\cite{Khatsymovsky:1998tv, Kuhfittig:2003wr, Kim:2004ph, Kashargin:2007mm, Bronnikov:2013zxa, Kleihaus:2014dla, Azreg-Ainou:2014aqa, Azreg-Ainou:2014nra, DelAguila:2015isj, Azreg-Ainou:2015pda, Chew:2016epf, Chew:2018vjp, Chew:2019lsa, Azreg-Ainou:2020ffg, Rahaman:2021web, Clement:2022pjr, Cisterna:2023uqf}.
To construct a rotating version of our charged wormhole,
we utilized the modified NJ algorithm, introducing $a$ rotation parameter.
Notably, the position of the throat of the rotating wormhole remains independent of any angular dependencies.

A notable reference solution for a rotating black hole is represented by the Kerr solution in a vacuum.
Since its inception, various geometries of rotating black holes that incorporate matter fields
or arise from modified gravity theories have been developed by extending the Kerr black hole geometry.
While it is essential to ensure consistency with the Einstein equations,
a well-established reference solution for the geometry of a rotating wormhole is still lacking.
Nonetheless, following Teo's initial discovery of the rotating wormhole geometry,
substantial research has been conducted in this area.
Given that we do not yet possess a satisfactory reference for rotating wormhole geometries,
it remains crucial to continue exploring their geometric properties
and the characteristics of the associated matter fields through diverse methodologies, including Azreg-Ainou's prescription~\cite{Azreg-Ainou:2015pda}.
Currently, there is no definitive algorithm for rotating geometries that satisfy the condition $-g_{tt}\neq g^{rr}$.
Consequently, we have slightly relaxed the mathematical rigor in our approach.
We have further advanced the NJ algorithm to extend its applicability to geometries possessing the attribute of $-g_{tt}\neq g^{rr}$.

The location of the rotating wormhole throat is where $\vartriangle_2$ goes to zero;
this location is greater than the location where $G(r, \theta)$ equals zero, thanks to the use of
Eqs.~(\ref{whrho}), (\ref{difwh1}), and (\ref{difwh02}) are designed to differ from their black hole counterparts.
It coincides with the locations where $G(r, \theta)$ is zero only at $\theta=0$ and $\pi$.
This characteristic may influence the metric function $dtd\phi$ in Eq.~(\ref{fullmetric}),
potentially leading to a singularity in the geometry at $\theta=0$ and $\pi$.
At this point, we have not been able to rigorously demonstrate and analyze the Kretschmann invariant in detail.
We will reserve a comprehensive analysis of this aspect for future research.

We have modified and applied the NJ algorithm to derive the rotating wormhole geometry.
However, we note that we have been unable to establish
that the $G_{r\theta}$ term in the Einstein equations is zero~\cite{Azreg-Ainou:2014aqa}.

We have analyzed the flare-out condition for this geometry to be a rotating wormhole.
For this purpose, we checked the geometry with ${\rm t=constant}$ and $\theta=\pi/2$,
and checked that the circumference radius has the diverging property at the throat of the wormhole.
A detailed analysis of the surplus angle will be addressed in future work.
Our observations establish constraints on the charge $Q$ and the parameter $a$.
We would like to emphasize that the rotation effect tends to increase the size of the rotating wormhole throat,
while the charge effect has the opposite influence, decreasing its size.
Additionally, we derived constraints on the values of the equation of state parameters
due to the rotational effects when $w_1 < -1$.

Finding a Maxwell tensor that satisfies the geometry of a rotating wormhole appears to be another challenging problem.
We will leave finding solutions to the Maxwell equations for future work.
When the metric function deviates from that of a rotating black hole~\cite{Kerr:1963ud, Newman:1965my, Toshmatov:2015npp, Kim:2019hfp, Clement:2023xvq},
the effects of this difference become evident in the Maxwell tensor\footnote{
We tried to find the solution by following the Refs.~\cite{Janis:1965tx, Newman:1965my, Stephani:2003tm, Kim:2021vlk}.
The procedure is for the rotating black holes~\cite{Kerr:1963ud, Newman:1965my, Toshmatov:2015npp, Kim:2019hfp},
where the metric function $f(r)$ is equal to $g(r)$ for the static ones.
However, we could not come to a successful close to obtaining solutions to Maxwell equations for the rotating wormhole geometry,
where $f(r)$ is not equal to $g(r)$ for the static ones.
Thus, we did not show the process in detail. One could see the modification of the Einstein-Maxwell equations~\cite{Cotton:2021tfl}.}.
In the Appendix, we present the components of the Maxwell tensor that satisfy the Maxwell equations
in the geometry of a charged rotating black hole.
This geometry corresponds to a situation in which the metric functions $f(r)$ and $g(r)$ are unequal.

When an electric field converges on a wormhole from one universe,
it passes through the wormhole throat and emerges into another universe.
At the throat of the wormhole, the Maxwell tensor goes to zero.
If we consider a Gaussian surface that encloses the asymptotic regions of both universes,
we observe that the flux of the electric field exiting is equal to that entering,
indicating that there is no net charge within the Gaussian surface.
We have demonstrated solutions to the source-free Maxwell equations by constructing a wormhole geometry with a charge $Q$.
We would like to emphasize that this geometry offers a geometric realization of the concept of ``charge without charge"~\cite{Misner:1957mt}.

We have obtained and analyzed wormhole solutions as a specific type of solution to general relativity.
However, our understanding of their geometric properties is not as advanced as it is for black hole solutions.
Even if the matter supporting a wormhole could possess negative density and radial pressure,
the wormhole's geometry could bend light to make a photon sphere--a circular orbit--and generate a strong gravitational field
with a capture cross-section. It could also exert tidal forces.
Much of the gravitational phenomena associated with the bending of light around a black hole
can be attributed to its stronger gravitational field,
with the Newtonian theory of gravitation serving as a reference for comparison.
However, challenges remain in interpreting and comprehending the geometrical properties of a wormhole geometry,
primarily because the Newtonian theory does not have this kind of solution.
We believe that sustained interest and research in wormhole physics will
enhance our understanding of the geometry of curved spacetime.

The investigation of wormhole stability presents an intriguing topic, which we will explore in future research~\cite{Kang:2018izw, Kim:2018ang, Kang:2019ost, Jonas:2023qle, Lee:2023jfi}.
The notion of charge without charge is naturally extended, thereby yielding the static traversable wormhole solutions
in the presence of a cosmological constant~\cite{Kim:2025zyo}.
The effects of magnetic fields are being observed and being analyzed in the shadows of the supermassive black hole in $M87$~\cite{EventHorizonTelescope:2025dua}.
It would also be worthwhile to conduct comparative studies investigating the effects of magnetic fields
on wormholes with an anisotropic matter field.

When you miss someone who has passed away, you may think about a wormhole or a time machine at least once.

\section*{Acknowledgments}
H.-C.~Kim (RS-2023-00208047), S.-W. Kim (2021R1I1A1A01056433), B.-H. Lee (RS-2024-00339204), and
W.~Lee (RS-2022-NR075087; CQUeST Grant: RS-2020-NR049598) were supported by Basic Science Research Program through
the National Research Foundation of Korea funded by the Ministry of Education.
We extend our gratitude to Jungjai Lee for his hospitality during our visit to the workshop on theoretical physics at Daejin University,
as well as to Wontae Kim and Stefano Scopel for their hospitality to the Chaiho Rim Memorial Workshop (CQUeST 2023),
and to Dong-han Yeom and Yun Soo Myung to SGC 2023.
We are grateful to Yun Soo Myung, Youngone Lee, Inyong Cho, Jae-Hyuk Oh for valuable discussions,
and Mustapha Azreg-Ainou for his helpful suggestions.
We appreciate the hospitality provided by APCTP during the completion of this work.

\vspace{1.5cm}

\appendix

{\large \bf Appendix}

\section{The geometry of a rotating black hole with the proper NJ algorithm}
\label{app:robh}

Here we describe the application of the Newman-Janis algorithm to the case of a rotating black hole.
It may be useful to compare this section with the rotating wormhole case in Sec.~[\ref{sec3}].

Instead of the Eq.~(\ref{whrho}), the following expression shall be used
\begin{equation}
{\bar\rho}^2_{\rm BH} = r^2 + a^2\cos^2\theta \,.
\label{bhrho}
\end{equation}
Here after we use $\rho^2$ instead of ${\bar\rho}^2_{\rm BH}$ for simplicity.
Instead of the Eq.~(\ref{tetrad4}), the null the tetrad becomes
\begin{eqnarray}
l_{\mu}&=& \delta^0_{\mu} - a\sin^2\theta \delta^3_{\mu}   \,, ~~
n_{\mu}=\frac{F(r, \theta)}{2}\delta^0_{\mu} + \sqrt{\frac{F(r, \theta)}{G(r, \theta)}}\delta^1_{\mu} + a\sin^2\theta \left( \sqrt{\frac{F(r, \theta)}{G(r, \theta)}} - \frac{F(r, \theta)}{2}\right) \delta^3_{\mu}   \nonumber \\
m_{\mu}&=&\frac{\rho^2}{\sqrt{2} (r+ia\cos\theta)}\left[\delta^2_{\mu} + i\sin\theta \delta^3_{\mu} \right]   \,, ~~
\bar{m}_{\mu}= \frac{\rho^2}{\sqrt{2} (r-ia\cos\theta)}\left[\delta^2_{\mu} - i\sin\theta \delta^3_{\mu} \right] \,.
\label{bhtetrad4}
\end{eqnarray}
Instead of the Eq.~(\ref{difwh1}), we use
\begin{eqnarray}
du=  dt - \frac{\sqrt{\frac{G(r, \theta)}{F(r, \theta)}}\rho^2 + a^2\sin^2\theta}{\triangle_{BH2}} dr \,, ~~~
d\psi= d\phi - \frac{a}{\triangle_{BH2}} dr \,,
\label{bhdifwh1}
\end{eqnarray}
to obtain Boyer-Lindquist coordinates
\begin{eqnarray}
ds^2&=& - F(r, \theta) dt^2 -2 \left(\sqrt{\frac{F(r, \theta)}{G(r, \theta)}} -F(r, \theta)\right) a\sin^2\theta dt d\phi + \frac{\Sigma}{\rho^2} \sin^2\theta d\phi^2
+ \frac{\rho^2}{\vartriangle_{BH2}} dr^2 + \rho^2 d\theta^2 \nonumber \\
&=& - \frac{\triangle_{BH3}}{\rho^2} (dt -a \sin^2\theta \triangle_{BH4} d\phi)^2 + \frac{\sin^2\theta}{\rho^2}[(r^2+a^2)\triangle_{BH5} d\phi -a dt]^2 + \frac{\rho^2}{\vartriangle_{BH2}} dr^2 + \rho^2 d\theta^2 \,,
\label{bhfullmetric}
\end{eqnarray}
where
\begin{eqnarray}
\Sigma & =&  (r^2+a^2)^2 - \vartriangle_{BH1} a^2 \sin^2\theta =  {\rho^2}\left(\rho^2 -a^2 \sin^2\theta \left(F(r, \theta) -2\sqrt{\frac{F(r, \theta)}{G(r, \theta)}} \right)\right) \,,
 \nonumber \\
\vartriangle_{BH1} & =& \rho^2 \left( F(r, \theta) -2 \sqrt{\frac{F(r, \theta)}{G(r, \theta)}} +2\right) + a^2 \sin^2\theta \,,  \nonumber \\
\vartriangle_{BH2} & =&\rho^2 G(r,\theta) + a^2 \sin^2\theta \,, ~~~~ \vartriangle_{BH4} =1-\sqrt{\frac{1}{F(r,\theta)G(r,\theta)}} \pm \sqrt{\frac{\vartriangle_{BH2}}{F(r,\theta)G(r,\theta)\vartriangle_{BH3}}}\,, \nonumber \\
\vartriangle_{BH3} &=&  \rho^2 F(r,\theta) + a^2 \sin^2\theta  \,, ~~~~ \vartriangle_{BH5} =\frac{a^2 \sin^2\theta \left(F(r,\theta) -\sqrt{\frac{F(r, \theta)}{G(r, \theta)}}  \right) \pm \sqrt{\frac{F(r, \theta)\vartriangle_3\vartriangle_2}{G(r, \theta)}}}{F(r,\theta)(r^2+a^2)} \,.
\label{bhdef}
\end{eqnarray}
If $F(r, \theta)=G(r, \theta)$, then $\triangle_{BH1}=\triangle_{BH2}=\triangle_{BH3}$, $\triangle_{BH4}=\triangle_{BH5}=1$
with taking the plus sign in $\triangle_{BH4}$ and the plus sign in $\triangle_{BH5}$,
thus all are reduced to the cases for the rotating black hole \cite{Kim:2019hfp}.

The metric functions $F(r, \theta)$ and $G(r, \theta)$ of the black hole, in Eq.~(\ref{bhdifwh1}),
have the event horizon at the same location, and that part is canceled by a common factor.
However, there is the remaining part.
Regardless of this one, we expect that Eq.~(\ref{bhfullmetric}) would be the geometry of a rotating black hole.
For instance, one can substitute the metric functions for the hairy black hole~\cite{Kanti:1995vq, Antoniou:2017acq, Doneva:2017bvd, Lee:2018zym, Lee:2021uis}
into $G(r, \theta)$ and $F(r, \theta)=G(r, \theta)e^{-\delta(r)}$, in which $\delta(r)$ goes to zero when approaching $r\rightarrow \infty$.
The location of the event horizon is determined from $\triangle_{BH2}=0$ and the location of the ergosphere from $F(r, \theta)=0$.

The inverse metric can be written as follows:
\begin{displaymath}
g^{\mu\nu} = \left(\begin{array}{cccc}
-\frac{G(r, \theta)[\rho^2 + a^2\sin^2\theta(2\sqrt{\frac{F(r, \theta)}{G(r, \theta)}}-F(r, \theta))]}{F(r, \theta) \triangle_{BH2}} & 0 & 0  & - \frac{G(r, \theta)a(\sqrt{\frac{F(r, \theta)}{G(r, \theta)}}-F(r, \theta))}{F(r, \theta) \triangle_{BH2}} \\
0 &  \frac{\triangle_{BH2}}{\rho^2} & 0 & 0 \\
0  & 0 & \frac{1}{\rho^2} & 0 \\
-\frac{G(r, \theta)a(\sqrt{\frac{F(r, \theta)}{G(r, \theta)}}-F(r, \theta))}{F(r, \theta) \triangle_{BH2}} & 0 & 0 & \frac{G(r, \theta)}{\triangle_{BH2}\sin^2\theta}
\end{array}  \right)\,,
\label{bhinversebl}
\end{displaymath}
where the determinant factor is
\begin{eqnarray}
\sqrt{-\det g_{\mu\nu}} = \rho^2 \sin\theta \sqrt{\frac{F(r,\theta)}{G(r,\theta)}} \,.
\label{bhdeterfac}
\end{eqnarray}
For Eq.~(\ref{bhfullmetric}) to be suitable for describing the geometry of a rotating black hole,
the $G_{r\theta}$ component of the energy-momentum tensor should be zero~\cite{Azreg-Ainou:2014aqa}.
If $F(r, \theta)=G(r, \theta)f(r)$, this fact leads to the following conditions:
\begin{eqnarray}
\label{TrthetBH}
&&a^2 \cos\theta \left[\rho^2 \sqrt{f(r)} \left( G(r, \theta)f'(r) + 2f(r) G'(r, \theta) \right) + 4f(r)(r-G(r, \theta))
-f'(r)(2\rho^2 +a^2\sin^2\theta) \right] \nonumber \\
&& + \rho^4 \left[ f(r){\dot G}'(r, \theta) +f'(r){\dot G}(r, \theta) \right]  = 0 \,,
\end{eqnarray}
where the overdot denotes the derivative with respect to $\theta$.

From Eq.~(\ref{bhfullmetric}), the set of covariant tetrad(co-tetrad) is as follows:
\begin{eqnarray}
&&e^{\hat t}_{\mu} = \frac{\sqrt{\triangle_{BH3}}}{\rho} (1,0,0,-a\sin^2\theta \triangle_{BH4})\,, ~~~
e^{\hat r}_{\mu} = \frac{\rho}{\sqrt{\triangle_{BH2}}} (0,1,0,0) \,, \nonumber \\
&& e^{\hat \theta}_{\mu} = \rho (0,0,1,0) \,, ~~~
e^{\hat \phi}_{\mu} =  \frac{\sin\theta}{\rho} (-a,0,0, (r^2+a^2)\triangle_{BH5})  \,.
\label{bhcotetrad}
\end{eqnarray}

{\large \bf Maxwell field}
\vspace{.3cm}

In this section, we present the solutions to the Maxwell equations.
When a charged black hole rotates, it can induce a magnetic field,
and this induced magnetic field and complicated geometry by the rotation make it difficult to solve
the Maxwell equations directly to find the solutions~\cite{Newman:1965my, Kim:2021vlk}.

We have experience in finding the Maxwell field for the charged rotating black hole with a matter field~\cite{Kim:2021vlk}.
Our objective is to apply this method directly to find the Maxwell field
for a charged rotating black hole geometry characterized by $-g_{tt}\neq g^{rr}$.
When solving Maxwell equations (\ref{maxwelleq}), to remove the functional dependence of the $\sqrt{-\det g_{\mu\nu}}$
on the function $\sqrt{\frac{F(r, \theta)}{G(r, \theta)}}$ shown
in Eq.\ (\ref{bhdeterfac}), we multiply the Maxwell field by $\sqrt{\frac{G(r, \theta)}{F(r, \theta)}}$,
which immediately satisfies Maxwell equations (\ref{maxwelleq}). They are as follows:
\begin{eqnarray}
F^{tr} &=& -F^{rt} = \sqrt{\frac{G(r,\theta)}{F(r,\theta)}} \frac{Q}{\rho^6} (r^2 - a^2\cos^2\theta)(r^2+a^2) \,,\nonumber \\
F^{t\theta} &=& -F^{\theta t} = \sqrt{\frac{G(r,\theta)}{F(r,\theta)}}\frac{Q}{\rho^6} (-a^2 r \sin2\theta)\,,  \nonumber \\
F^{r\phi} & = & - F^{\phi r} = \sqrt{\frac{G(r,\theta)}{F(r,\theta)}} \frac{Q}{\rho^6} a (a^2\cos^2\theta - r^2) \,,\nonumber \\
F^{\theta\phi} &=& -F^{\phi\theta} = \sqrt{\frac{G(r,\theta)}{F(r,\theta)}}\frac{Q}{\rho^6} 2ar\cot\theta \,.
\label{bhmaxcou}
\end{eqnarray}
For the Kerr-Newman type rotating black hole, $F(r,\theta)=G(r,\theta)$ as shown in \cite{Kim:2021vlk}.

In the asymptotic rest frame with $r \gg a$, the non-vanishing electromagnetic fields with (\ref{bhcotetrad}) takes the usual form
\begin{eqnarray}
E^{\hat r} &=& F^{{\hat t}{\hat r}}  \simeq \frac{Q}{r^2}  \,, \nonumber \\
B^{\hat r} &=& F^{{\hat \theta}{\hat \phi}} \simeq 2 \frac{Q a}{r^3} \cos\theta \,, \nonumber \\
B^{\hat \theta} &=& F^{{\hat \phi}{\hat r}} \simeq \frac{Q a}{r^3} \sin\theta \,.
\label{bhexpaelecmag}
\end{eqnarray}
The above are the induced dipole magnetic field and ${\cal M}=Q a$ corresponds to the magnetic moment of the wormhole.
They are the same for the black hole~\cite{Misner:1973prb}.

\end{document}